\documentclass[aps,prstper,twocolumn,showpacs,superscriptaddress,letterpaper]{revtex4-1}

\usepackage{graphicx}
\usepackage{subfigure}
\usepackage{color}
\usepackage{enumitem}
\usepackage{mdframed}
\usepackage{framed}
\usepackage{epstopdf}
\usepackage{amsmath}
\newcommand{\textactivityone}{
\begin{itemize}\footnotesize
\item[a)]{Build a voltage divider similar to the one shown in Fig. \ref{fig:act1} using resistors of around $1~\mathrm{k\Omega}$. Draw a diagram of the circuit in your lab book. Make sure to label the resistors and record all measured component values and voltages.}
\item[b)]{Measure each resistor with your DMM before inserting it into your circuit and record the value. Why should you measure component values before placing them in the circuit?}
\item[c)]{Predict the output voltage you should measure based on your input voltage and resistance measurements. Include your calculations and numerical predictions in your lab book.}
\item[d)]{Now, apply a DC voltage to the input and measure the output voltage of your divider, first using your DMM and second using your oscilloscope with the mini-grabbers. Record your measurements. (Do not have the DMM and oscilloscope connected at the same time because each may perturb the measurement differently.)}
\item[e)]{Compare the voltages you predicted to the voltages you measured. Does your model of the voltage divider agree with each of your measurements? Explicitly record what criteria you used to determine whether or not the model and measurements agree.}
\item[f)]{\emph{Complete this step only if your model and measurements did not agree.} If your model and measurements did not agree, you will have to either refine your model or your experiment. Let's start by refining your model. Consider the input resistance of your measurement device. Draw a circuit diagram that includes that resistance. \emph{HINT: You already worked with this circuit model in your pre-lab.} Derive an expression for the output voltage now including the unknown measurement device resistance. Use this mew model to determine the input resistance of the measurement device.}
\item[g)]{Complete steps {\bf a-f} for two additional voltage dividers, one using resistors $\sim1~\mathrm{M\Omega}$ and $\sim10~\mathrm{M\Omega}$}
\item[h)]{Using your refined model, you have determined the input resistance of both the DMM and scope. Specification sheets or data sheets can also be used to refine your model.}
\end{itemize}\normalsize
}

\newcommand{\textactivitytwo}{
\begin{itemize}\footnotesize
\item[]{\underline{Lab prep activities}}
\item[a)]{What intensity of white light in $\mathrm{mW/cm^2}$ do you expect is incident on your photodiode on the lab bench when it is facing upwards, i.e. towards the fluorescent lights? Each fluorescent light tube produces approximately $4\mathrm{W}$ of visible light. You can assume that half of it is emitted downwards into $2\pi$ sr.}
\item[b)]{Be sure to state your assumptions explicitly for part (a). How many bulbs did you model and at what distance from the detector was each bulb? What wavelength are you assuming? Etc. State at least three key assumptions in your model prediction.}
\item[]{\underline{Photometer}}
\item[a)]{Measure the average intensity of light from the fluorescent lamps in the lab from the output of your photometer circuit [shown in Fig. \ref{fig:act2}]. How does your result compare with your lab prep estimate? Keep in mind that the estimate you made of the light intensity was very rough, and also note that the data sheet only gives a ``typical'' value of the sensitivity of the photodiode. How could you refine your model to more accurately represent your measurement system or your physical system to more accurately represent your model? List two possible refinements and complete at least one. Explain why you think this refinement could allow you to get better model-measurement agreement. Report on what you did, your new measurements/model predictions, and if you were successful in getting better agreement.}
\end{itemize}\normalsize
}

\newcommand{\textactivitythree}{
\begin{itemize}\footnotesize
\item[a)]{Build the circuit for [the voltage-controlled electromagnet], using your coil.}
\item[b)]{Using the data you acquired [previously], determine $V_{supply}$ and $V_{GS}$ to operate your electromagnet with nearly 1A of current in the saturated regime. Remember to include the voltage drop across your coil in your calculations (HINT: $V_{DS}=V_{supply}-I_D R_{coil}$ in your model).}
\item[c)]{Test out your electromagnet. Choose a way to test your model of the magnetic field produced. Explain the model you are testing and what your predictions are (even qualitative, such as how things scale).}
\item[d)]{Describe your procedure, results/measurements, and refinements to your model. Some starter suggestions are listed below. Be creative! 
\begin{itemize}
\item{Measure the magnetic field as a function of distance from the plane of the loop, using a Gaussmeter.}
\item{Look at the force applied by a permanent magnet on the coil and see how that scales with current or number of turns.}
\item{Check the effect of your coil on a compass and compare to the known magnetic field of the Earth.}
\item{Get a power supply that can supply more current and pick up objects, test how the number or weight of objects you can pick up depends on the current.}
\item{Wrap your coil around an iron core and test how the magnetic field changes, refine your model to include the iron.}
\end{itemize}}
\end{itemize}\normalsize
}

\newcommand{\figureactivityone}{
\begin{figure}[h]
\centering 
\includegraphics[scale=0.35]{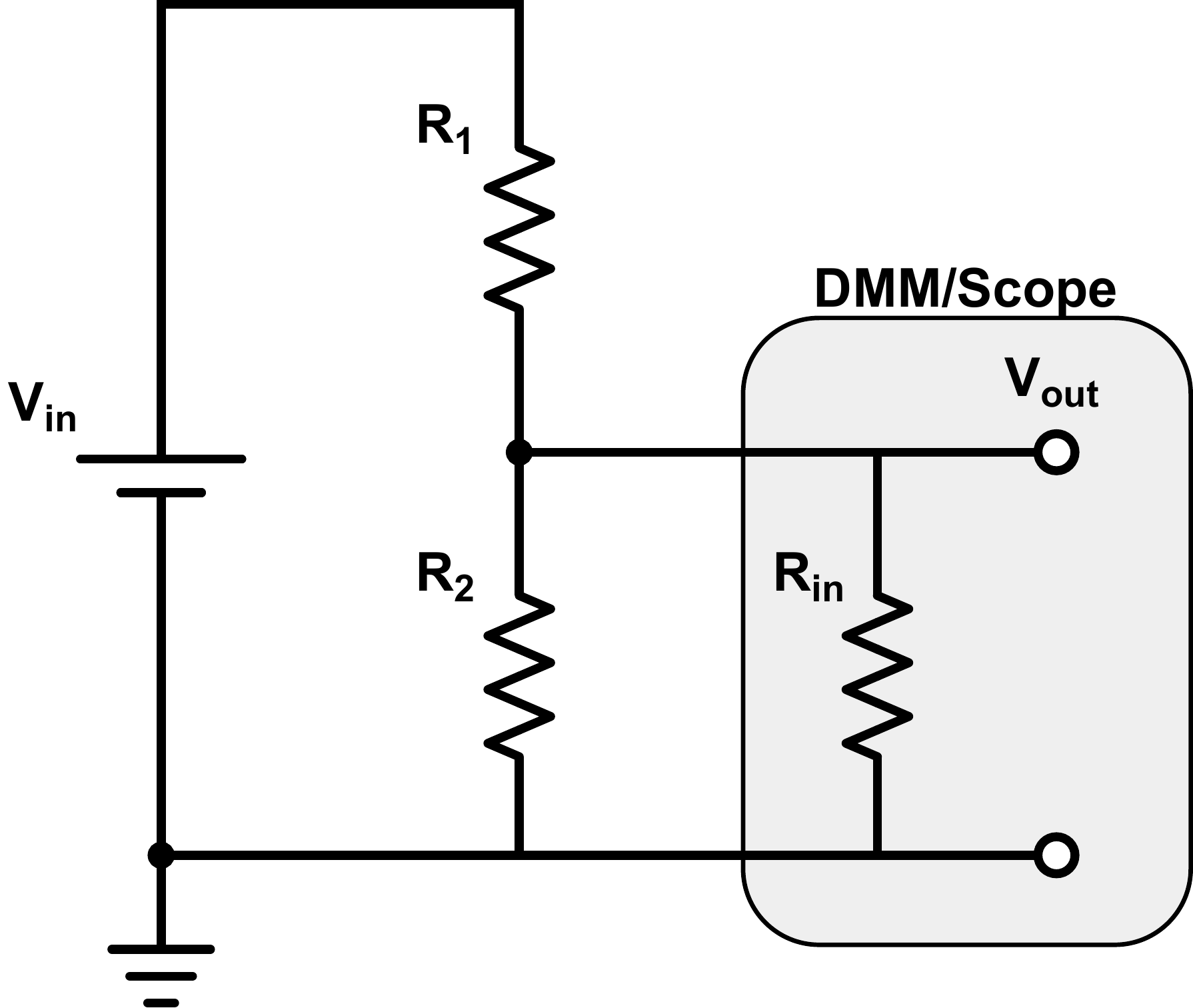}
\caption{Voltage divider circuit for Activity 1 (a version of this, without the measurement device shown, was included in the students' pre-lab). Resistors $R_1$ and $R_2$ were both either $1\,\mathrm{M\Omega}$ or $10\,\mathrm{M\Omega}$, depending on the version of the circuit being tested. The measurement device is depicted with the input resistance labeled $R_{in}$. Everything in the box is internal to the measurement device.}
\label{fig:act1}
\end{figure}
}
\newcommand{\figureactivitytwo}{
\begin{figure}[h]
\centering 
\includegraphics[scale=0.5]{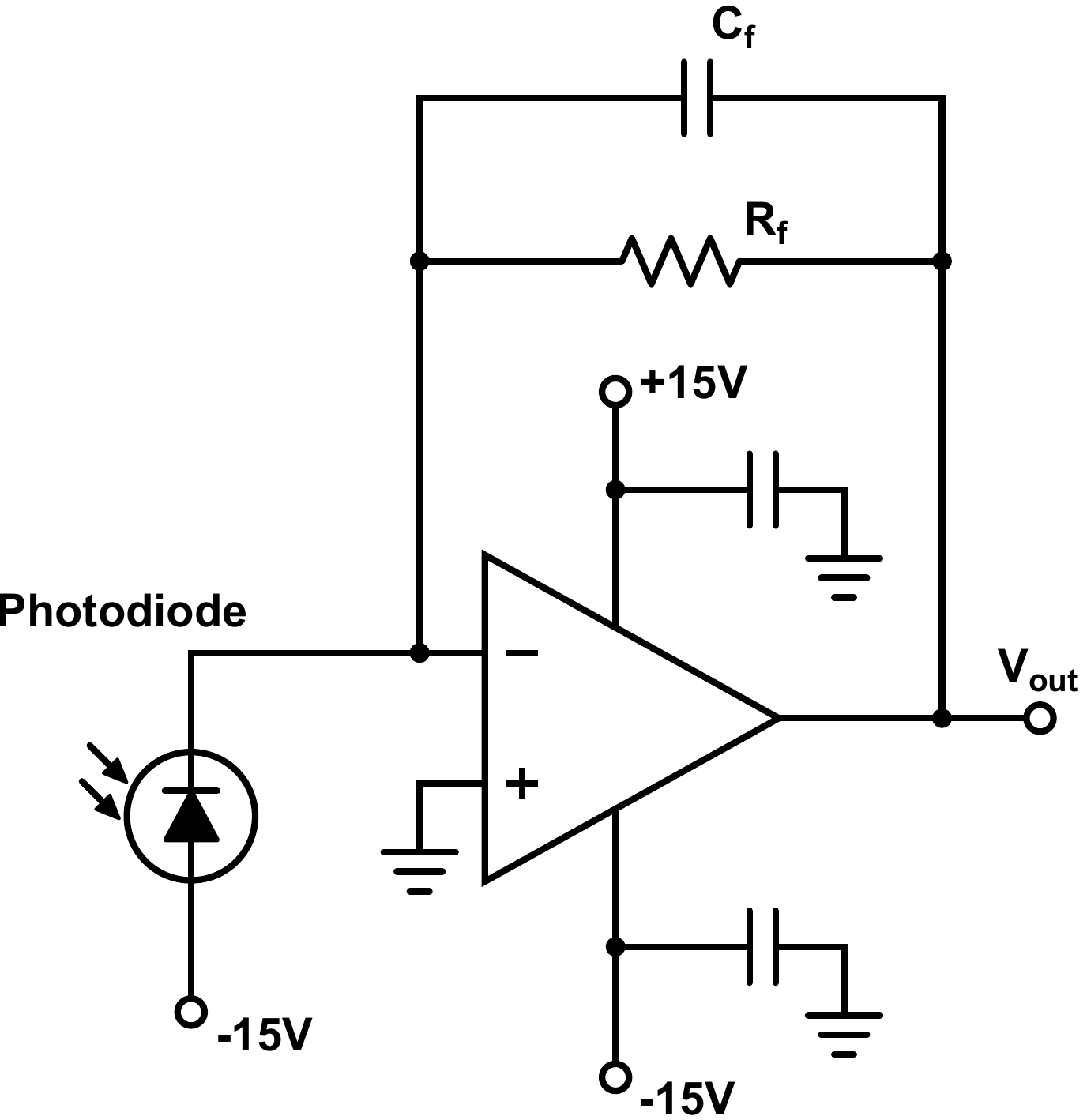}
\caption{Photometer circuit for Activity 2. The photodiode students used to detect the room light is visible at the bottom left and is in a reverse-biased configuration. The op-amp portion of the circuit is set up as a transimpedance amplifier.}
\label{fig:act2}
\end{figure}
}
\newcommand{\figureactivitythree}{
\begin{figure}[h]
\centering 
\includegraphics[scale=0.45]{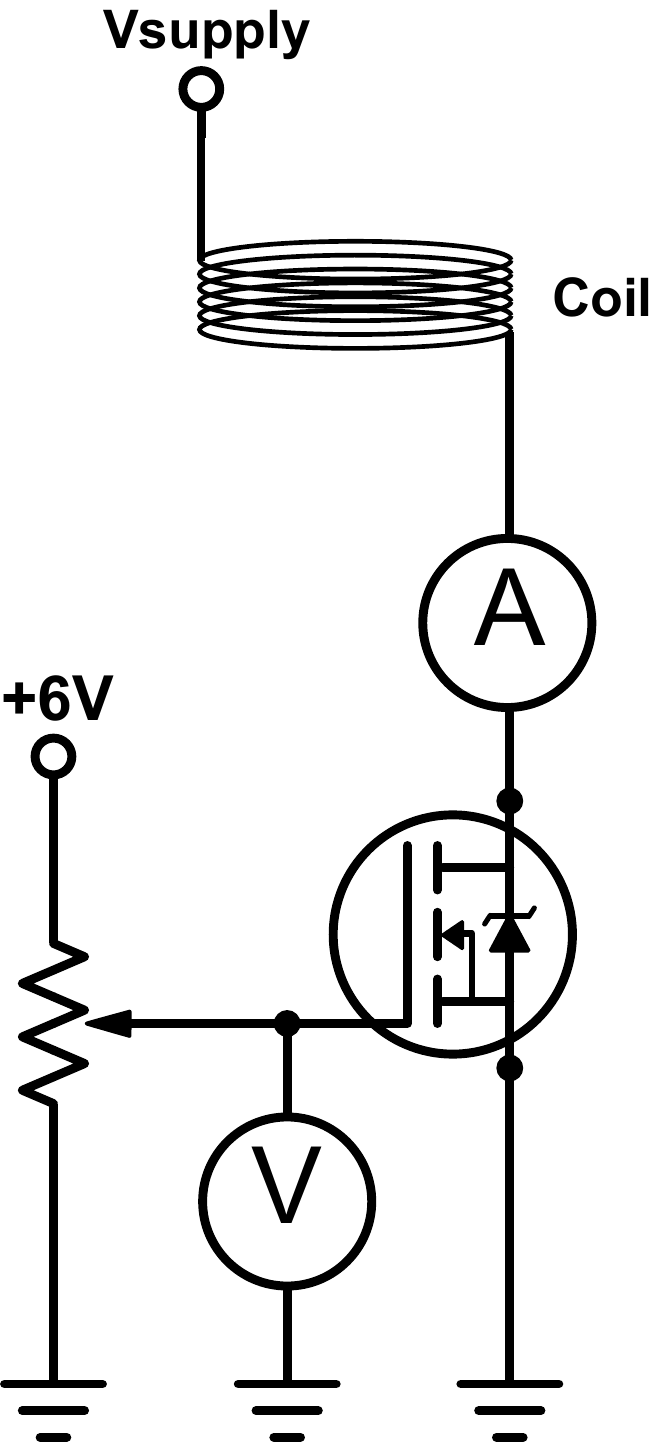}
\caption{Voltage-controlled electromagnet circuit used for Activity 3. The component labeled ``A'' is the ammeter used to measure the current through the coil and the component labeled ``V'' is the voltmeter used to measure the gate to source voltage.}
\label{fig:act3}
\end{figure}
}
\newcommand{\actonecoderes}{
\begin{figure*}[t]
\centering 
\includegraphics[scale=0.85]{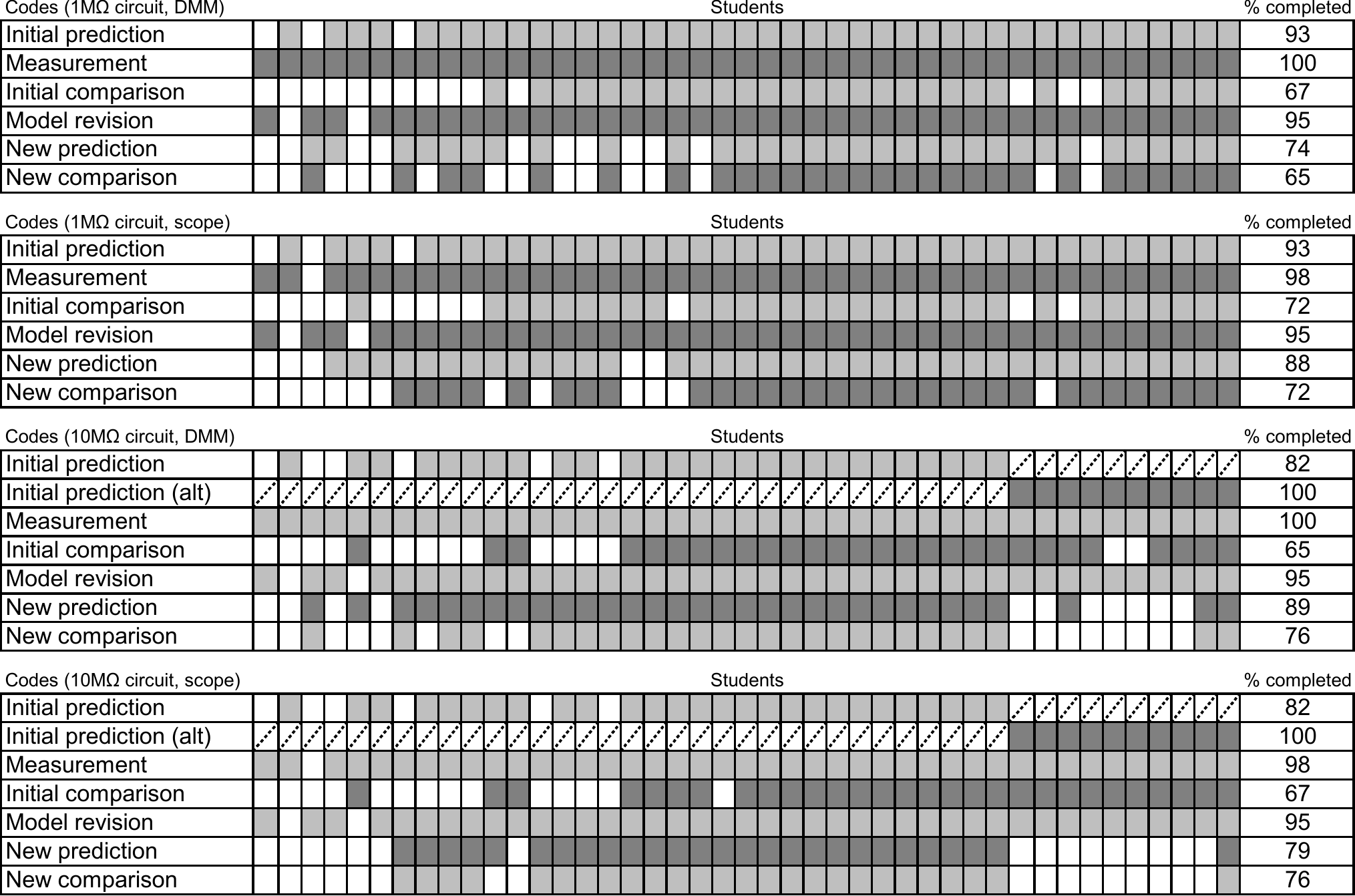}
\caption{Coding results for the four parts of Activity 1. The top (bottom) two diagrams show the coding results for the $1\,\mathrm{M\Omega}$ ($10\,\mathrm{M\Omega}$) circuit measured with the DMM. The results for the oscilloscope measurements of the same circuits demonstrated essentially the same degree of completion. In total, 37\% of the students documented the modeling process for the entire activity (all four parts). Rows represent different steps in the modeling process and columns represent individual students (the corresponding column of each of the four diagrams represents the same student). Each grey box represents a documented instance of the corresponding modeling step for the corresponding student. The diagonal hash marks indicate that the two different initial predictions represented two separate modeling pathways and thus no student would do both. Right most column represents percentage of students who documented that step of the modeling process. The percentages for the initial predictions are based on the total number of students who chose one of the two modeling pathways.}
\label{fig:resactivity1}
\end{figure*}
}
\newcommand{\acttwocoderes}{
\begin{figure*}[t]
\centering 
\includegraphics[scale=0.9]{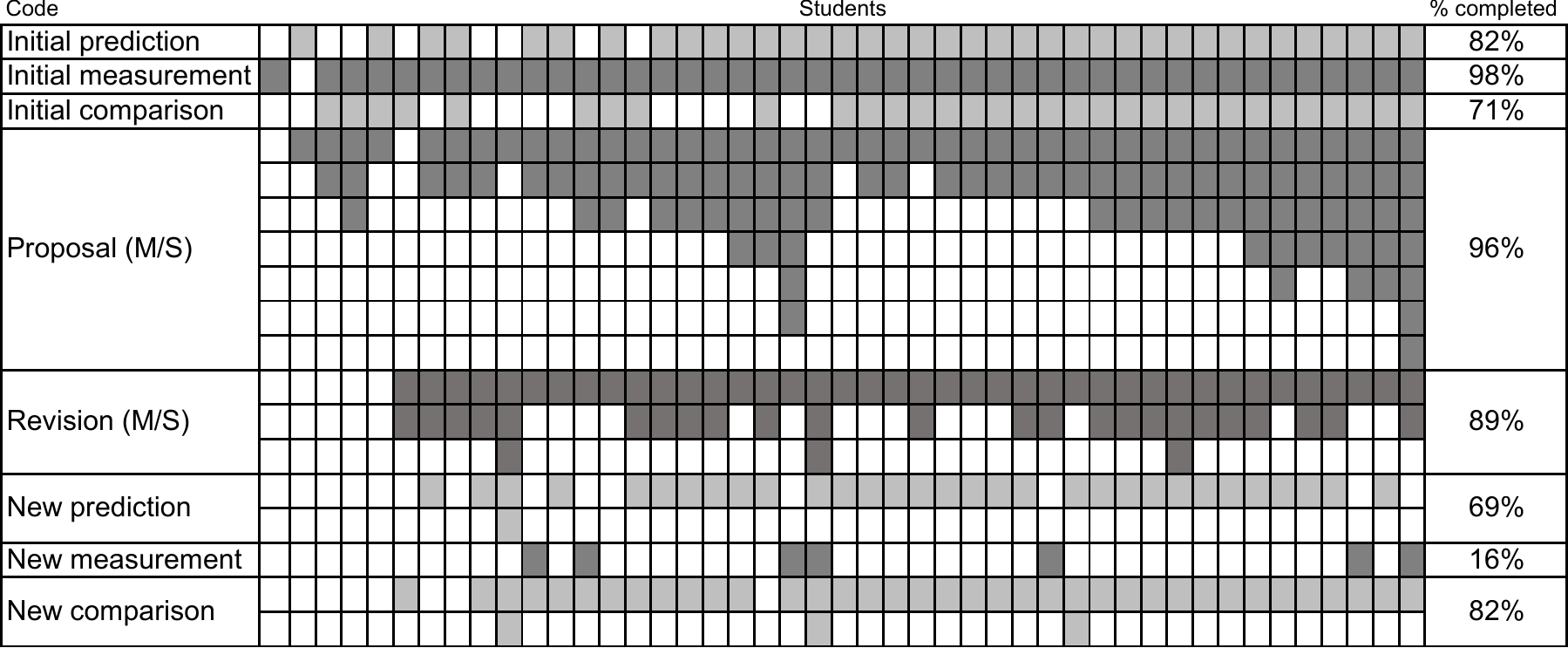}
\caption{Coding results for Activity 2. In total, 53\% of the students completed the modeling process for the activity. Rows represent different steps in the modeling process and columns represent individual students. Each grey box represents a documented instance of the corresponding modeling step for the corresponding student.}
\label{fig:resactivity2}
\end{figure*}
}
\newcommand{\actthreecoderes}{
\begin{figure*}[t]
\centering 
\includegraphics[scale=0.85]{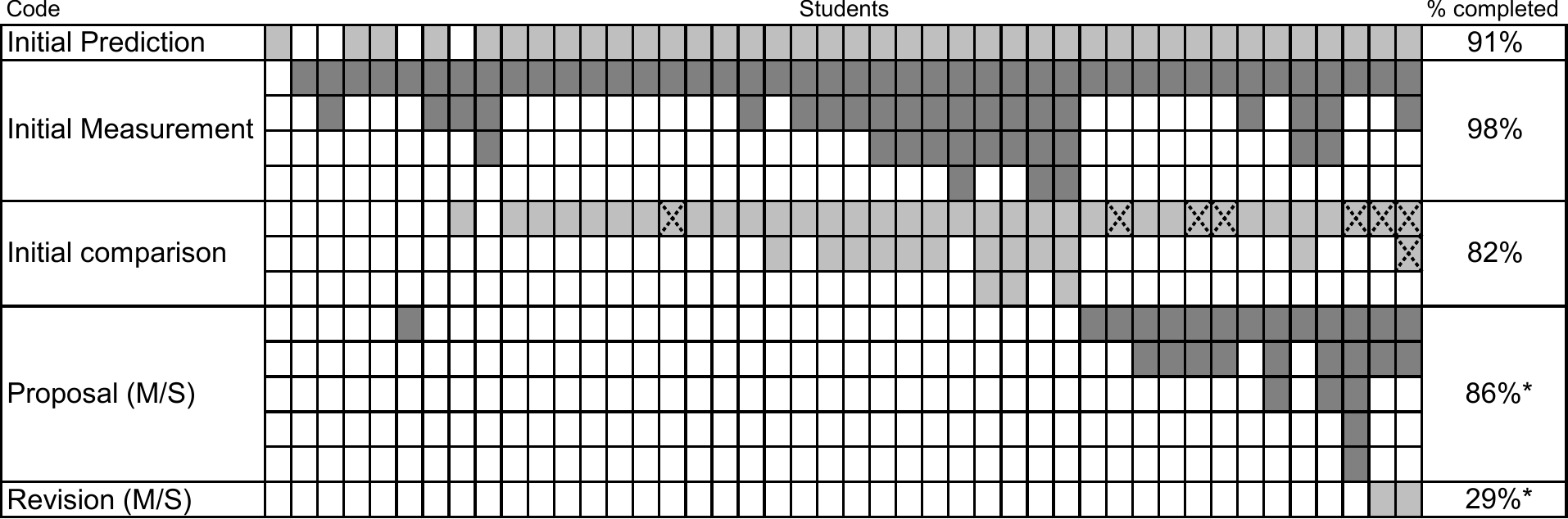}
\caption{Coding results for Activity 3. In total, 68\% of the students completed the modeling process for the activity. Rows represent different steps in the modeling process and columns represent individual students. Each grey box represents a documented instance of the corresponding modeling step for the corresponding student. The X's indicate comparisons that were evaluated to be in poor agreement. Percentages (*) for proposals and revisions were based on the total number of students whose comparisons indicated poor agreement.}
\label{fig:resactivity3}
\end{figure*}
}
\newcommand{\actonecompex}{
\begin{figure}[t]
\centering 
\includegraphics[scale=0.42]{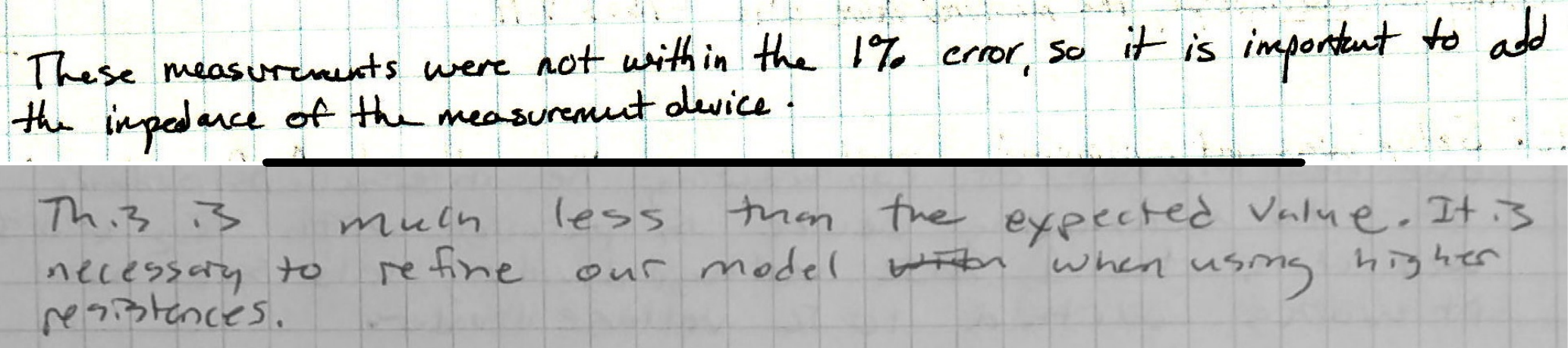}
\caption{Two examples of documented comparisons for Activity 1. The top example demonstrates the type of quantitative reasoning students used, while the bottom demonstrates the type qualitative reasoning. Note that a proposal to refine the system is also stated in both examples.}
\label{fig:compact1}
\end{figure}
}
\newcommand{\acttwocompex}{
\begin{figure}[h]
\centering 
\includegraphics[scale=0.42]{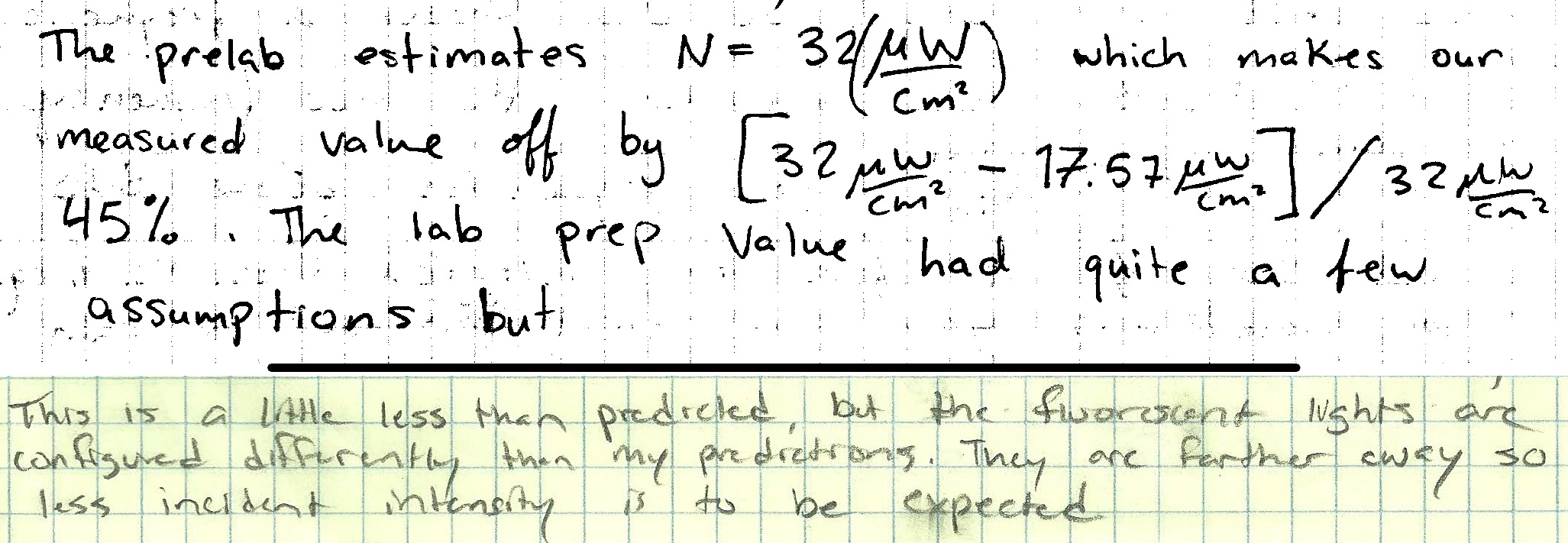}
\caption{Two examples of documented comparisons for Activity 2. The top example demonstrates a quantitative comparisons using percentage difference. The bottom example makes a subjective qualitative statement (``a little less than predicted'') as a comparison.}
\label{fig:compact2}
\end{figure}
}
\newcommand{\actthreecompex}{
\begin{figure}[h]
\centering 
\includegraphics[scale=0.42]{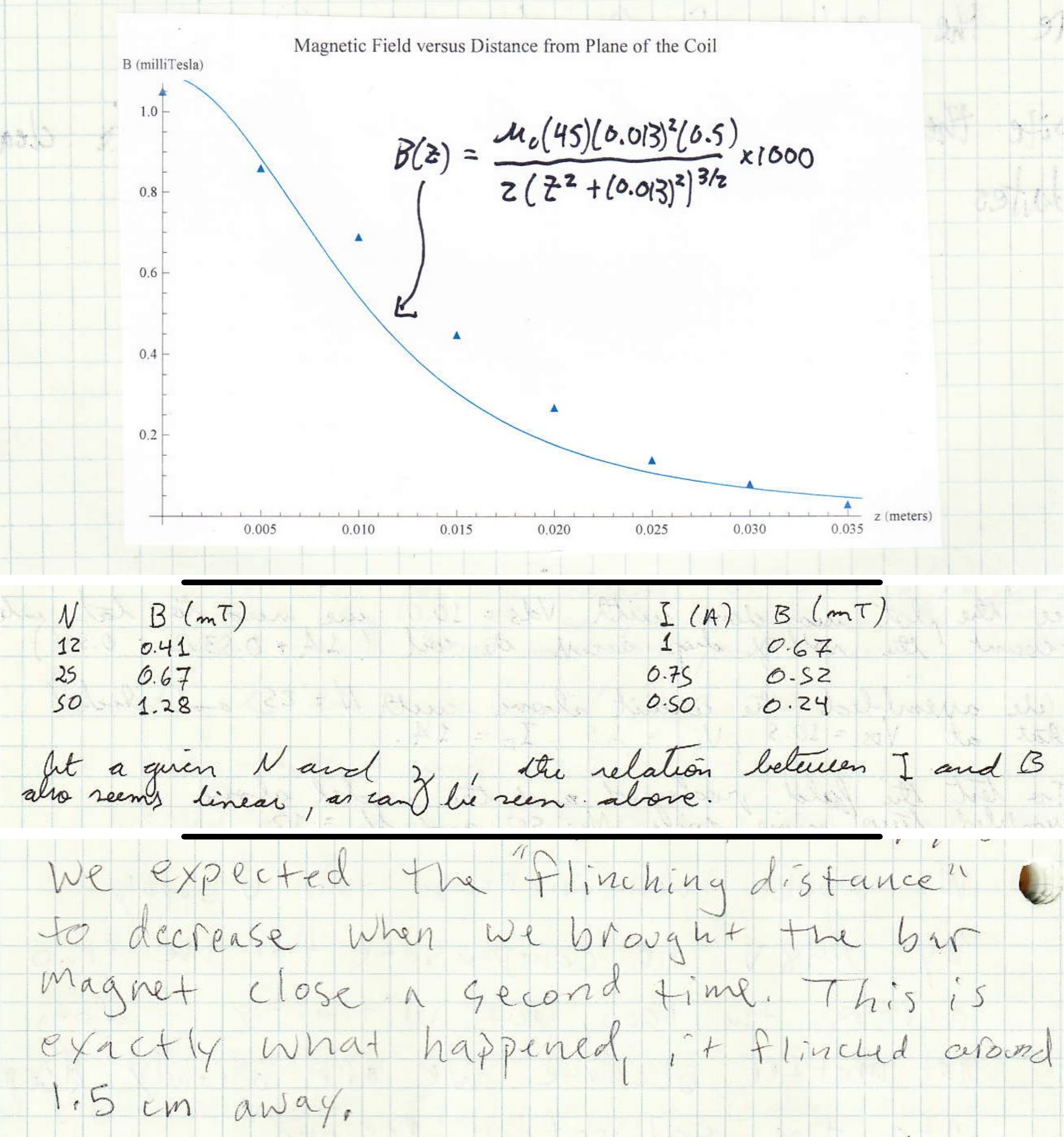}
\caption{Three examples of documented comparisons for Activity 3. The top example demonstrates the type of comparison plots students made for a model that directly measured the B-field. The middle example demonstrates a proportionality comparison to their model, $B \propto N$ and $B \propto I$. The bottom example demonstrates a trend comparison where the model addresses only the increase or decrease of the B-field.}
\label{fig:compact3}
\end{figure}
}
\newcommand{\activitycodes}{
\begin{table*}[h!]
\caption{Finalized consensus codes for all three activities that were the result of coding process described in Sec. \ref{sec:codeprocess}. The Activity 1 codes were used for both the 1 and 10$\mathrm{M\Omega}$ resistor voltage divider circuits. Note the similarity of the codes for each activity---all three activities included prediction, measurement, comparison, and revision codes. Since the revision pathway in Activity 1 was constrained by the lab guide, there exists only one revision code (\emph{i.e.}, model revision) and no codes for proposals. The code sets for Activities 2 and 3 contain proposal codes, as well as revision codes for multiple pathways due to their open-ended nature. These proposal and revision codes are the only codes in this table that have associated sub-codes describing the specific types of proposals and revisions students performed (discussed in Sec. \ref{sec:resact2} and \ref{sec:resact3}).}
\begin{ruledtabular}
\begin{tabular}{p{3.6cm} p{13.6cm}}
\emph{Code name} & \emph{Definition}\\ \hline
\emph{   Activity 1} & \\
\rule{0pt}{3.5ex}Initial prediction& Student predicts the output voltage of the voltage divider ($V_{out}$) using the resistance values of the two resistors ($R_1$ and $R_2$), assuming infinite input resistance for the measurement device.\\
\rule{0pt}{3.5ex}Initial prediction (alt)& An alternate prediction where the student predicts the output voltage of the voltage divider ($V_{out}$) using the two resistor values ($R_1$ and $R_2$) and the input resistance of the measurement device, $R_{in}$ (obtained from previous measurement or data sheet)\\
\rule{0pt}{3.5ex}Measurement& Student makes initial measurement of $V_{out}$ with the measurement device (digital multimeter or oscilloscope)\\ 
\rule{0pt}{3.5ex}Initial comparison& Student compares the initial or alternative prediction to the initial measurement.\\ 
\rule{0pt}{3.5ex}Model revision& Student constructs a model of the circuit that includes the resistance of the measurement device. Consists of a circuit diagram and an equation for $V_{out}$.\\
\rule{0pt}{3.5ex}New prediction& Student makes a new prediction of $V_{out}$, calculated with the input resistance of the measurement device as a parameter obtained from spec sheet or other source OR student makes a prediction of the input resistance of the measurement device, calculated using their initial voltage measurement for $V_{out}$ in their model.\\
\rule{0pt}{3.5ex}New comparison& Student compares their new prediction of $V_{out}$ with their initial measurement OR student compares their prediction of the input resistance of the measurement device to a reference obtained from data sheet or other source.\\ \hline
\emph{   Activity 2} & \\
\rule{0pt}{3.5ex}Initial prediction& Student makes prediction of the room light intensity.\\
\rule{0pt}{3.5ex}New prediction& Student updates their prediction of the room light intensity from that made in the pre-lab (may occur prior to taking an initial measurement).\\
\rule{0pt}{3.5ex}Initial measurement& Student perform a measurement of the room light intensity with their initial equipment set-up (must be in the same units as the prediction).\\ 
\rule{0pt}{3.5ex}New measurement& Student perform a measurement of the room light intensity after having performed a revision of the system.\\ 
\rule{0pt}{3.5ex}Initial comparison& Student compares their prediction to their initial measurement.\\
\rule{0pt}{3.5ex}New comparison& Student makes any kind of comparison between a new prediction and a new measurement that resulted from a revision to either the model or the system.\\
\rule{0pt}{3.5ex}Proposal (model)& Student proposes a revision to the model anticipated to improve agreement between the subsequent new prediction and their measurement.\\
\rule{0pt}{3.5ex}Proposal (system)& Student proposes a revision to the physical system anticipated to improve agreement between the subsequent new measurement and their prediction.\\
\rule{0pt}{3.5ex}Revision (model)& Student performs a revision to their model.\\
\rule{0pt}{3.5ex}Revision (system)& Student performs a revision to their physical system.\\ \hline
\emph{   Activity 3} & \\
\rule{0pt}{3.5ex}Initial prediction& Student provides a qualitative or quantitative prediction about the behavior of the magnetic field.\\
\rule{0pt}{3.5ex}Initial measurement& Student makes a qualitative or quantitative measurement of the B-field.\\
\rule{0pt}{3.5ex}Initial comparison& Student compares their initial measurement to the initial prediction.\\ 
\rule{0pt}{3.5ex}Proposal (model/system)& Student proposes a revision to either the system or their model.\\ 
\rule{0pt}{3.5ex}Revision (model/system)& Student performs a revision to their system or model.\\
\end{tabular}
\end{ruledtabular}
\label{tab:codes}
\end{table*}}

\begin{document}

\title{Using lab notebooks to examine students' engagement in modeling in an upper-division electronics lab course}

\author{Jacob T. Stanley}
\affiliation{Department of Physics, University of Colorado, Boulder, CO 80309, USA}
\author{Weifeng Su}
\affiliation{Department of Physics, University of Colorado, Boulder, CO 80309, USA}
\affiliation{Department of Physics, Fudan University, Shanghai 200433, China}
\author{H. J. Lewandowski}
\affiliation{Department of Physics, University of Colorado, Boulder, CO 80309, USA}
\affiliation{JILA, National Institute of Standards and Technology and University of Colorado, Boulder, CO, 80309, USA}

\begin{abstract}
We demonstrate how students' use of modeling can be examined and assessed using student notebooks collected from an upper-division electronics lab course. The use of models is a ubiquitous practice in undergraduate physics education, but the process of constructing, testing, and refining these models is much less common. We focus our attention on a lab course that has been transformed to engage students in this modeling process during lab activities. The design of the lab activities was guided by a framework that captures the different components of model-based reasoning, called the Modeling Framework for Experimental Physics. We demonstrate how this framework can be used to assess students' written work and to identify how students' model-based reasoning differed from activity to activity. Broadly speaking, we were able to identify the different steps of students' model-based reasoning and assess the completeness of their reasoning. Varying degrees of scaffolding present across the activities had an impact on how thoroughly students would engage in the full modeling process, with more scaffolded activities resulting in more thorough engagement with the process. Finally, we identified that the step in the process with which students had the most difficulty was the comparison between their interpreted data and their model prediction. Students did not use sufficiently sophisticated criteria in evaluating such comparisons, which had the effect of halting the modeling process. This may indicate that in order to engage students further in using model-based reasoning during lab activities, the instructor needs to provide further scaffolding for how students make these types of experimental comparisons. This is an important design consideration for other such courses attempting to incorporate modeling as a learning goal.
\end{abstract}

\maketitle
\section{\label{sec:introduction}Introduction}

Constructing, testing, and refining models of the physical world is a core practice in physics research, and constitutes the crux of the \emph{process of modeling}.\cite{Etkina2006,Schwarz2009} In physics, a model is an abstract representation of a physical phenomena that, in part, serves to simplify, encode, and communicate the essential features of that phenomenon. Commonly, these models are represented by equations, diagrams, words, and graphs, which facilitate the formation of testable predictions germane to the phenomenon being modeled.\cite{Zwickl2015,Zwickl2014,Schwarz2009,Hestenes1987}

The importance of incorporating instruction on the process of modeling into science education has been recognized across the physical sciences, and at most levels of education.\cite{NGSS2013,NRC2012,NRC1998} For undergraduate physics lab courses in particular, the American Association of Physics Teachers Committee on Laboratories has released guidelines that emphasize modeling as one of the six major learning outcomes for laboratory courses.\cite{AmericanAssociationofPhysicsTeachers2014}

Though physics students constantly work with models in both lecture and lab courses, the process of modeling is not often explicitly addressed in the undergraduate curriculum. There has been increasing effort by the education research community to improve understanding of the process modeling and how to implement it in physics education, but much of this effort has focused primarily on the lecture course environment---not the instructional lab environment.\cite{Brewe2008, Etkina2006, Halloun1996, Wells1995, Hestenes1987}

To lay the groundwork for incorporating modeling into laboratory courses, a framework has recently been developed that specifically addresses the experimental physics environment---the Modeling Framework for Experimental Physics (EMF), which can be seen in Fig. \ref{fig:emf}.\cite{Zwickl2015,Zwickl2014,Zwickl2013} Previous work has used this framework to inform the transformation of two upper-division physics lab courses, focusing on electronics and modern physics, at the University of Colorado Boulder.\cite{Lewandowski2015,Zwickl2015} In addition to the EMF being used to guide the design of lab course activities, it was intended be used to guide systematic observation and evaluation of students' model-based reasoning, either in real-time or in the analysis of students' written materials.

To date, this framework has been used to study model-based reasoning in experimental physics activities during think-aloud interviews---video recordings of students verbalizing their in-the-moment thinking while engaging in a physics activity.\cite{Zwickl2015} Specifically, the EMF mapped well onto students' performance on optics and electronics physics tasks, by capturing students' modeling process during these experimental activities.\cite{Dounas-Frazer2016,Dounas-Frazer2015,Zwickl2015} Although these efforts have demonstrated the applicability of the EMF to real time student reasoning, it has (i) not yet been utilized to probe model-based reasoning in a lab course setting and (ii) it has not been used to extract model-based reasoning from written data sources such as student lab notebooks. The work herein is the first to address both these avenues. Specifically, instead of examining a structured experimental physics activity outside of a course environment, we will use the EMF to examine students' model-based reasoning in an upper-division electronics laboratory course. Also, instead of using video as our data source, we will analyze the written documentation in students' lab notebooks for model-based reasoning.

First and foremost, the goal of this work is to illustrate how the EMF can be used to analyze and evaluate model-based reasoning in lab notebooks by examining the extent to which students explicitly document their modeling process. The lab course we focus on is the newly transformed electronics lab course at the University of Colorado Boulder. Two of the primary learning goals for this lab course are (i) to engage students in modeling during lab activities and (ii) to have students practice authentic scientific documentation in lab notebooks. Thus, this course is an ideal environment for this study. In contrast to previous efforts\cite{Zwickl2015}, we are using the EMF as a tool to assess student work. In this paper, we present the results of analyzing three different lab activities that demonstrate varying degrees of activity scaffolding and potential for modeling. In our analysis, we compare and contrast the specifics of the modeling students documented during these lab activities.

A secondary outcome of our efforts is to provide a preliminary evaluation of how well this transformed course is engaging students in the practice of modeling. In our analysis of these activities, we answer the following questions regarding the model-based reasoning students demonstrated: (i) Are students documenting recursive modeling cycles in their notebooks? (ii) Do differing levels of scaffolding in lab activities influence students' documented modeling cycles? If so, how? (iii) With which components of the modeling process do students demonstrate more or less proficiency? The answers to these questions will help to provide insight into how the transformed course can be improved to better promote students' engagement with, and adoption of, modeling practices. In turn, these have implications for lab course design more broadly.

\section{\label{sec:framework}Theoretical framework}

\begin{figure}[h]
\centering 
\includegraphics[scale=0.75]{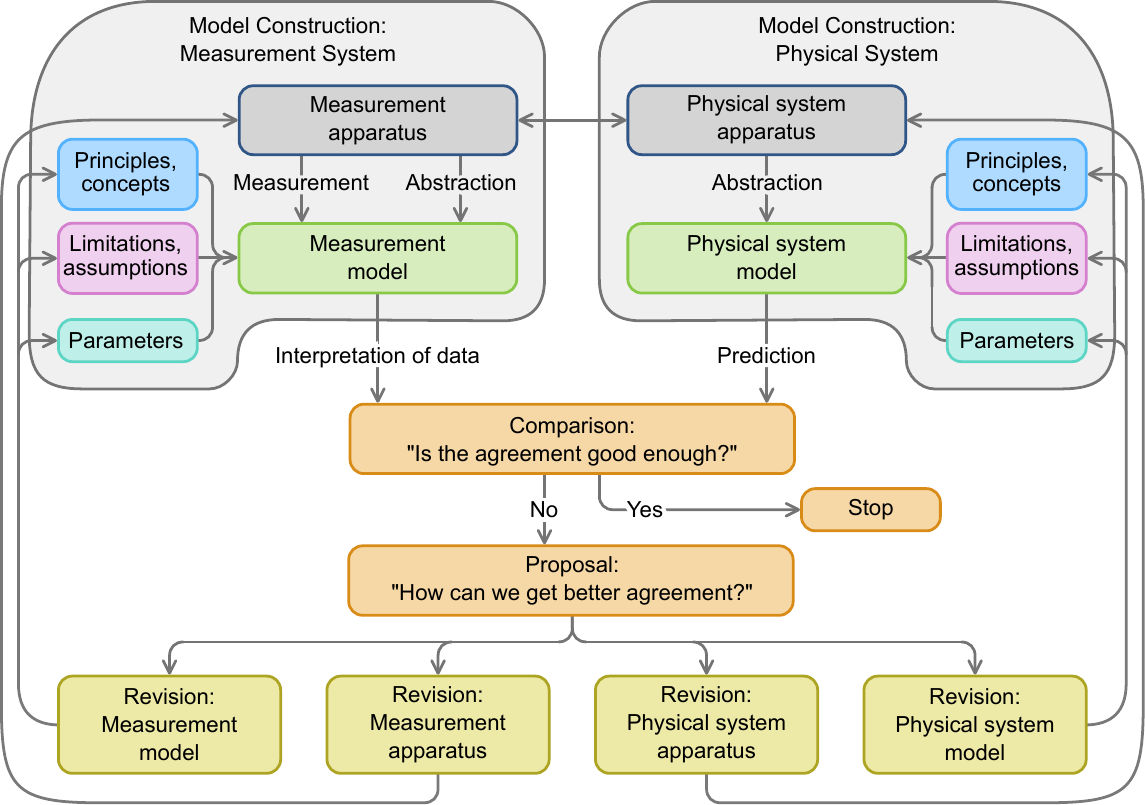}
\caption{Diagram for the experimental modeling framework.}
\label{fig:emf}
\end{figure}

The theoretical approach to this project is based on a previously developed framework designed to be applied to the physics laboratory course environment.\cite{Zwickl2014,Zwickl2015} A diagram of this framework is depicted in Fig. \ref{fig:emf}. One of the major features of this framework is consideration for the measurement apparatus, in addition to the physical system. This is depicted in the diagram by showing the full experimental apparatus broken down into two main components: the physical system (right side of Fig. \ref{fig:emf}) and the measurement system (left side of Fig. \ref{fig:emf}). Both sub-systems are given equal representation.

Due to the complexity of measurement apparatus for many upper-division labs, one can not understand the physical system being studied without also having an understanding of how the measurement apparatus functions. Furthermore, modeling of the measurement apparatus is an essential part of lab activities in an authentic research setting. Thus, incorporating practice with this aspect of modeling will benefit students going on to participate in authentic research.

In creating the framework, it was acknowledged that the division between these two sub-systems is not always unique---in many cases there may be multiple ways of conceptualizing the division between the two. However, any reasonable division between the two will benefit the modeling process. The left-right symmetry of the framework's diagram emphasizes that, regardless of where the division is made, the physical and measurement systems both must be modeled.

As described by the framework, the modeling process starts with the initial \emph{model construction} (depicted in the top left and top right of Fig. \ref{fig:emf} for the measurement and physical systems, respectively), which consists of incorporating the \emph{principles}, \emph{concepts}, \emph{limitations/assumptions}, and key \emph{parameters} into the abstraction of the apparatus. Commonly, the construction of the physical system model is based on the physics concepts learned in undergraduate courses, while the construction of the measurement system model is based on the technical documentation for the equipment.

The models for the two sub-systems are then used to make a \emph{prediction} (using the physical system model) and \emph{interpret the raw data from a measurement} (using the measurement system model).

The next step in the modeling process is to \emph{compare} the prediction and interpreted data. The comparison is evaluated to determine if there is sufficient agreement between the prediction and measurement to stop the modeling process (the ``comparison'' bubble in Fig. \ref{fig:emf}). If there is not sufficient agreement, the next step in the modeling process is to \emph{propose a revision} to the model or system.

The framework identifies four distinct revision pathways (depicted at the bottom of Fig. \ref{fig:emf}): revise the model describing the measurement apparatus, revise the measurement apparatus itself, revise the physical system apparatus, or revise the model describing the physical system. Depending on which revision has been completed, one then proceeds to perform a new measurement or formulate a new prediction. This process forms a cycle, which is repeated until one has achieved adequate agreement between the model prediction and the data. 

It should be noted that what constitutes ``adequate'' agreement is very context dependent. In some cases, an order of magnitude comparison is sufficient where as other circumstances might require high precision. Thus, there is no single criteria one can use to determine when to stop the process. Careful consideration for what comparison criteria to use is an essential part of the process of modeling.

\section{\label{sec:transformation}Overview of course transformation}

The work presented here was in the context of a recently transformed junior-level electronics lab course taught in the physics department at the University of Colorado (CU). This required course for physics and engineering physics majors covers mostly analog electronics with a small component on digital electronics. The course structure includes 20 one-hour lectures on concepts relevant for the lab activities. The lab component consists of 10 one-week guided lab activities and a five-week-long student inspired project. Each lab section meets once a week for three hours, however, the students have 24/7 swipe card access to the lab room to be able to finish the activities. 

	The transformation of this course is part of a larger effort to improve education in experimental physics throughout the curriculum at CU. The goals for the transformation effort were developed using the broader learning goals previously identified by the faculty\cite{Zwickl2013} and through discussions with faculty that regularly teach the course. The desired student outcomes include expertise with (1) using measurement and design equipment (oscilloscopes, prototyping boards, DMMs, etc.), (2) proper data collection and measurement techniques, and (3) characterizing, modeling, and understanding applications of  core components (discrete components, voltage dividers, operational amplifiers, transistors, etc.). In addition, faculty wanted to see increased student satisfaction with, and engagement in, the course, and have activities represent authentic practice of experimentalists who work with electronics. 
	
	It is equally important to note what were not learning goals for the course. Formal propagation of errors and error analysis are not goals for the course. The students are required to take two prerequisite lab courses that focus almost exclusively on error propagation and analysis. We did not have students go through formal error analysis for the labs in the transformed course, as we found that it was dominating the students time and cognitive resources, thus not allowing them to achieve the other learning goals. We also did not concentrate on the solid-state physics concepts that underpin the function of the electronic components. Most of the students in the class are taking the first semester of quantum mechanics concurrently with the lab course, and therefore do not have the background to understand band theory, Fermi levels, etc. This is not an overly restrictive constraint on the class, since these concepts are not required (even by experts) to design, build, troubleshoot, and use analog circuits. 

To work towards meeting these transformation goals, several major changes were made to the course. Most importantly for this work, the lab guides and prelab questions were re-written to engage students in the process of modeling. Students were also introduced to the modeling framework (Fig. \ref{fig:emf}) in lecture and in discussions in the lab. A large poster-sized version of the framework hangs on the wall of the lab, and is used to discuss the components of the lab activities. In addition, formal lab reports were removed from the course. Students now record their measurements, models, and in-the-moment thinking in a lab notebook, which is then used for grading. 

	These changes, in addition to other modifications, were introduced and refined over one year. Essentially all of the changes have been sustained over the last three years by nine different faculty members teaching the course who were not directly involved with the transformation.

\section{\label{sec:methods}Methods}

In this section, we provide a description of our methodological approach to this study. We outline the details of the data sources and participants, as well as the coding process for each activity.

\subsection{Participants and data source}

This study centered on the junior-level electronics laboratory course offered by the physics department at the University of Colorado Boulder---a large, predominantly white, public university with highest research activity and a large physics program. Typical enrollment for the course ranged from 25 to 55 students per semester. The student demographics for the course, over the period 2005--2014 (a total of 725 students who completed the course) were as follows: men 86\% and women 14\%; White 76\%, Asian 7\%, Underrepresented Minorities 6\%, and Other/Unknown 11\%.

This study included a total of 45 students across three sections (17, 14, and 14 students, respectively) of the transformed electronics lab described in Section \ref{sec:transformation}. Each section was taught by a separate instructor (one of whom is an author on this paper), but all three instructors were aware of, or involved with, the course transformation. All sections were run in the same manner (same lab guides, course framing, lectures, etc.).
 
The data for this study were scans of student notebooks, which were collected and graded as a normal part of the course. Students signed consent forms that granted permission to use these materials for research purposes and did not receive any form of compensation. We focused on all content from the notebook that were directly associated with the lab activities of interest (\emph{i.e.}, the text, calculations, graphs, plots, and tables recorded by the student).

\subsection{\label{sec:codeprocess}Coding process}

The research team collaboratively identified a number of activities in the lab manual of the transformed course that prompted students to engage in modeling. We considered activities that had the potential for at most two or three cycles of the modeling process. This set of activities was reduced down to three that demonstrated varying degrees of scaffolding (the activities differed in how explicit were the modeling instructions). The research team determined that together the three activities mapped onto most of the components of the framework. These three activities were analyzed for the work herein. 

A separate coding scheme was developed for each lab activity. The coding scheme for each activity was based on the EMF and was developed in the following steps: develop the preliminary coding scheme, perform preliminary coding pass and creation of emergent sub-codes, consolidate the sub-codes into a broad consensus coding scheme, apply the consensus coding scheme to the data, and reconcile any discrepancies between coders in the final coding. Next, we describe these steps in more detail.

First, for each activity, all three authors discussed which components of the EMF were anticipated to be present in the students' notebooks, and through this discussion the authors established an \emph{a priori} preliminary coding scheme. This preliminary coding scheme helped focus the initial coding pass on the portions of the students' notebooks that were germane to modeling. The preliminary coding scheme was applied independently by two of the authors to the notebook data. During the preliminary coding, emergent codes were added to capture the detailed elements/features of student reasoning that were not captured by the preliminary coding scheme.

Second, the results of the preliminary coding were then discussed by all three authors. Each code was examined to determine if it captured the anticipated aspect of modeling, whether it should be consolidated with other codes, or if it should be eliminated from the coding set. The outcome of this discussion was a refined scheme of broad consensus codes that corresponded to the main components of the modeling framework (\emph{e.g.}, prediction, comparison, revision, as depicted in Fig. \ref{fig:emf}). Due to the open-ended nature of some of the activities, there existed a number of sub-codes for the proposal and revision codes. The sub-codes were detailed codes that identified the specific ways in which students engaged in the corresponding modeling step. Since our main focus was on the broad modeling process, we have not presented all the sub-codes here. However, the sub-codes did help us to organize the qualitative discussion about the different proposals and revisions students performed. We present examples of these sub-codes as a part of this discussion in Sec. \ref{sec:results}.

All codes were binary---they indicated the presence or absence of the particular modeling step. This consensus coding scheme (presented in Table \ref{tab:codes}) was used for the remainder of the coding/analysis, the results of which are presented in Figs. \ref{fig:resactivity1}, \ref{fig:resactivity2}, and \ref{fig:resactivity3}.

The consensus coding scheme for each activity was applied independently to the notebook data by two of the authors. Then, the coding by the two authors was compared, and each discrepancy was discussed and reconciled.

Finally, we collectively determined what constituted a complete modeling process independently for each activity, which we based on the finalized consensus coding scheme. What constituted a complete modeling process is discussed in the results in Sec. \ref{sec:results} for all three activities.

\subsection{Qualitative results and notebook examples}

The coding quantitatively captured the students' overarching modeling process. However, the coding does not illuminate the details of how the students engage in the individual components of the process. Thus, in addition to the quantitative coding (described in Sec. \ref{sec:codeprocess}), we also present some qualitative details about how students engaged in each step in all three activities. These qualitative results help provide context for the coding results and give one a better understanding of student reasoning. We identified the most illustrative and prominent details for each step by reading through the coded instances and selecting those that were the most common. For the parts of the modeling process for which we had sub-codes (\emph{e.g.}, proposals and revisions to the model/system), we identified the sub-codes with the highest frequency and provided a qualitative description of those. For the comparison step, we also present excerpts of student notebooks for each activity so one can better understand the nature of the comparisons being made, which is integral to understanding students' overall engagement in modeling.

\section{\label{sec:activities}Lab activities}
 
In this section, we describe the three activities selected for coding: (A) the resistive voltage divider, which was highly scaffolded, (B) the room light photometer, which was moderately scaffolded, and (C) the voltage-controlled electromagnet, which was the least scaffolded activity. Also, so that one may better understand the context of the coding, we have provided the wording of the activity prompts taken from the lab guides. Lastly, we present the final set of codes for each activity and provide justification for how the codes were decided upon.

\subsection{Activity 1: Resistive voltage divider}

In the first activity, students built a voltage divider with two different sets of resistors (two $1~\mathrm{M\Omega}$ and two $10~\mathrm{M\Omega}$ resistors, respectively) and measured the output voltage with two measurement devices, a digital multimeter (DMM) and an oscilloscope. The students initially modeled the output voltage using Ohm's Law with the assumption that the internal resistance of the measurement device was infinite ($R_{in}=\infty$ in Fig. \ref{fig:act1}). The model they established for the output voltage of the circuit ($V_{out}$) is represented by Eq. \ref{eq:vout1}.
\begin{equation}\label{eq:vout1}
V_{out}=\frac{R_2}{R_1 + R_2}\cdot V_{in}
\end{equation}
Upon measuring the output voltage with both measurement devices, they found that the measurements did not agree with their prediction due to the finite internal resistance of the measurement device. At this point, the students incorporated the finite input resistance of the measurement device into their diagrammatic and mathematical models of the system. The updated mathematical model was then represented by Eq. \ref{eq:vout1}.
%
\begin{align}\label{eq:vout2}
V_{out}=\frac{R_{eq}}{R_1 + R_{eq}}\,V_{in}\,\,\,\,\,&\mathrm{where}&R_{eq}=\frac{R_{2}R_{in}}{R_2 + R_{in}}&
\end{align}
Using their new model, the students either used a data sheet to obtain a value for $R_{in}$ and generated a new prediction for $V_{out}$ or they used their initial measurement of $V_{out}$ to make a prediction of $R_{in}$ (solving Eq. \ref{eq:vout2} for $R_{in}$), for each measurement device. In this way, the students generated a model, made a measurement and a prediction using that model, compared these two, and iterated on the model to generate a new prediction. This process constitutes a single cycle of modeling. 

This activity was highly scaffolded, in that students were explicitly prompted at each step of the process, and the potential modeling decisions were highly constrained (i.e., the model, measurement, and proposal/revision, were all specified). This activity was a part of the second lab of the course. The details of the lab guide for this activity are as follows: 
\textactivityone
\figureactivityone

The final set of codes for Activity 1 can be found in Table \ref{tab:codes}. Since students made measurements with two different measurement devices (DMM and oscilloscope) on two different circuits, they potentially performed four iterations of modeling that were related to one another. Therefore, the codes were applicable to the students' documentation for all four portions of the activity. The students worked on the $1$ and $10\mathrm{M\Omega}$ resistor voltage dividers in sequence. Thus, many students carried over information from one measurement to the next (namely, the known internal resistance of the measurement devices), where as other students made a prediction for the second circuit that was directly analogous to that made for the first circuit. Therefore, two different prediction codes were needed (\emph{initial prediction} and \emph{initial prediction (alt)}) to capture this difference in student reasoning. Furthermore, when refining their model and making a new prediction, some students used the measured voltage to predict the measurement device input resistance while others used the measurement device input resistance to predict/verify $V_{out}$. The \emph{new prediction} code captures both these paths. We evaluated all these different pathways as being acceptable model-based approaches to this activity.
%

\subsection{Activity 2: Room light photometer}

In the second activity, students utilized a transimpedance amplifier with a photodiode as a photometer to measure the average intensity of room light. During the pre-lab activities, students were prompted to make a prediction of the room light that would be detected by their photometer based on assumptions about the photodiode sensitivity, source-to-detector distance, and the theoretical intensity of the individual light bulbs in the room. The mathematical component of the model the students used was given by Eq. \ref{eq:act2},
\begin{align}\label{eq:act2}
&V_{out}=GS_{\lambda}\frac{nP}{2\pi R^2}
\end{align}
in which $n$ and $P$ represent the number and power of the light bulbs (respectively) and $R$ represents the source-to-detector distance. Also, $S_{\lambda}$ represents the sensitivity of the photodiode at a given wavelength of light and $G$ represents the gain of the photometer circuit (Fig. \ref{fig:act2}). Lastly, $V_{out}$ was the voltage students measured from the photometer circuit. This mathematical model incorporated parameters corresponding to both the physical system ($n$, $P$, and $R$) and the measurement system ($G$ and $S_{\lambda}$).

In lab, students were explicitly instructed to measure the ambient room light ($V_{out}$) at their lab bench and to compare this with their pre-lab prediction. Students were then asked to make refinements to either the model or the physical system and to justify their choice of refinements (unlike in Activity 1, they were not instructed to perform a specific refinement). They were then prompted to make new measurements or predictions, and to again compare their results.

The modeling in this activity was both less scaffolded and less constrained than that of the first activity. Furthermore, multiple refinement pathways were possible and students were expected to choose their own. This activity was a part of the sixth lab of the course. The details of the lab guide for this activity are as follows:
\textactivitytwo
\figureactivitytwo

The final set of codes for Activity 2 can be found in Table \ref{tab:codes}. Like Activity 1, there was a specific prediction and measurement that students were prompted to make in Activity 2 (the ambient room light intensity). Unlike Activity 1, there was a wide range of ways that students could revise the system or model in order to improve the agreement between their prediction and their measurement. We coded for students' proposed revisions. We did this because students were prompted to make a number of proposals and then select one to perform, so the proposals were an integral part of the model-based reasoning we wanted to capture. Furthermore, we coded for the specifics of the proposed and performed revisions, but given that our interest was in the general components of the modeling process (those depicted in the diagram of our theoretical framework, Fig. \ref{fig:emf}), these specific revisions were consolidated into broader codes that distinguish only between the physical system and the model. These broader codes are the last four in the Activity 2 section of Table \ref{tab:codes}.
%

\subsection{\label{sec:act3}Activity 3: Voltage-controlled electromagnet}

In the third activity, students built a voltage-controlled electromagnet using a coiled wire and a MOSFET. The goal of the activity was to test some aspect of the magnetic field produced by the the coil. Students were provided with a number of different examples of what could be tested (seen below in the text of the activity). Students were then instructed to develop a model that encapsulated this feature of the magnetic field, take measurements (either qualitatively or quantitatively), compare their measurement to their model's prediction, and perform revisions/refinements if needed. 

The modeling in this activity was less scaffolded and less constrained than either of the two previous activities. Students were given no specifics about any of the steps of the modeling process. Furthermore, there were no explicit restrictions on the models of the magnetic field the students could test. This activity was part of the eighth lab of the course. The details of the lab guide for this activity are as follows:
\textactivitythree
\figureactivitythree

The final set of codes for Activity 3 can be found in Table \ref{tab:codes}. Since there was little constraint to the predictions and revisions students could make regarding the magnetic field, we coded for the range of specific predictions and revisions students made, but ultimately consolidated those into broader codes, which captured only the major components of the modeling framework (codes 1, 4, and 5 in the Activity 3 section of Table \ref{tab:codes}). Furthermore, we did not include a \emph{``new measurement''} or \emph{``new prediction''} code like those in the previous two activities because in Activity 3 no students performed these steps.
\activitycodes

\section{\label{sec:results}Results}

In this section, we present the results of our analysis for each of the three activities outlined in Sec. \ref{sec:activities}. Our results include the outcome of the coding process described in Sec. \ref{sec:methods}, which captured the students' overarching modeling process. However, while our coding scheme gives us a clear picture of which components students engaged in, it does not give us insight into how they engaged in those components. Therefore, to provide a more complete picture, we also discuss qualitative examples of the types of reasoning students utilized in each of the activities.

\subsection{Activity 1: Resistive voltage divider}
\actonecoderes

Activity 1 consisted of four similar modeling cycles, one for each of the four parts of the activity: $1\,\mathrm{M\Omega}$ circuit measured with DMM, $1\,\mathrm{M\Omega}$ circuit measured with oscilloscope, $10\,\mathrm{M\Omega}$ circuit measured with DMM, and $10\,\mathrm{M\Omega}$ circuit measured with oscilloscope. The coding results for all four parts of Activity 1 can be seen in Fig. \ref{fig:resactivity1}. Each row corresponds to the steps in the modeling process that are represented by the codes in Table \ref{tab:codes}. The percentage of students who documented each of those steps is listed in the far right column of the figure. Each column represents an individual student and the shaded boxes indicate which components of the modeling process they documented having completed. The corresponding columns for each of the four parts of Fig. \ref{fig:resactivity1} correspond to the same student.

For each part of the activity, there were two different processes of reasoning that we considered to be a complete modeling cycle. One consisted of an initial prediction, measurement, comparison, new model construction (that incorporated the finite input resistance of the measurement device), a new prediction, and a new comparison. The other process consisted of an alternate initial prediction (which already accounted for the finite input resistance), measurement, and comparison. If the student documented either of these, they were considered to have completed the modeling process for that part of the activity, since both approaches ultimately resulted in the same revised model. Figure \ref{fig:resactivity1} is organized by these two different paths, as well as how complete was the students modeling process throughout the entire activity (left to right, from least to most complete). 

Overall, $74\%$ of students completed at least one of the four modeling cycles, $65\%$ completed at least two, $51\%$ completed at least three, and $37\%$ completed all four modeling cycles in the activity. Corresponding to Fig. \ref{fig:resactivity1} (from top to bottom), $51\%$ of the students completed the first modeling cycle, $58\%$ completed the second, $58\%$ completed the third, and $61\%$ completed the fourth modeling cycle of the activity.

The initial prediction, measurement, and model revision were the most commonly documented components of the modeling process for each part of the activity. Conversely, the initial comparison and new comparison were the least commonly documented components.

For the initial prediction, students only had to calculate $V_{out}$ using Eq. \ref{eq:vout1}. This model was established prior to the lab section as a part of their pre-lab assignment. Despite this, a number of students did not document a prediction (three for the $1\,\mathrm{M\Omega}$ circuit and six for the $10\,\mathrm{M\Omega}$ circuit). These students who did not document initial predictions also subsequently did not document their initial comparisons.

For the comparisons, there were 15 students for the $1\,\mathrm{M\Omega}$ circuit and 16 for the $10\,\mathrm{M\Omega}$ circuits who did not document it. This included all those that did not document their prediction, given that a prediction was prerequisite to the comparison. It is unclear whether or not students were actually performing the comparisons, but most of those students that did not document it still proceeded on to the later steps of the activity. This suggests they were performing the comparison, but did not recognize that it was important to document it. An alternative interpretation is that they felt the comparison was not necessary, despite being prompted to perform it.

The comparison was between a quantitative measurement ($V_{out}$) and the prediction generated from the model of their voltage divider. Regardless of whether or not the documented comparison indicated close enough agreement, the students proceeded with revising their model (as prompted by the lab guide). Predominantly, students made two types of comparisons---one quantitative and one qualitative:
\begin{itemize}
\item{Quantitative: Students made an evaluation of the comparison based on a percentage error criteria that was not previously established or referenced in the class.}
\item{Qualitative: Students made a subjective  comparison based on some unspecified criteria. These evaluations were qualitative in nature, along the lines of ``the measurement is reasonably close to the prediction'' or ``these measurements are in poor agreement with the model.''}
\end{itemize}
Examples of these two types of comparisons can be seen in Fig. \ref{fig:compact1}
\actonecompex

In addition to the specific modeling steps, an interesting result of Activity 1 was that despite the high degree of scaffolding students found more than one modeling pathway to follow. Students were documented using an alternate initial prediction (used for the $10\,\mathrm{M\Omega}$ circuit), in which they carried over the finite input resistance values of their measurement devices determined from measuring the $1\,\mathrm{M\Omega}$ circuit. In this event, they would start with the revised model (Eq. \ref{eq:vout2}) and thus did not need to do any further revision upon having made the initial comparison. They did this despite the fact that the activity prompt instructed students to ``predict the output voltage ... based on [their] input voltage and resistance measurements'' (step C in the Activity 1 text) for each of the four parts without incorporating the finite input resistance. These students recognized that that they could utilize the updated model that incorporated the finite input resistance of their measurement devices. In this manner, the students took a different modeling path by starting at a different point in the modeling framework. Of the 43 students, 10 of them took this alternative path (seen on the right side of Fig. \ref{fig:resactivity1}).

\subsection{\label{sec:resact2}Activity 2: Room light photometer}
\acttwocoderes

Activity 2 consisted of a single modeling cycle, which had a number of potential revisions to the model or system that could improve agreement between the model and measurement. The coding results for Activity 2 can be seen in Figure \ref{fig:resactivity2}. A complete modeling cycle consisted of an initial prediction, initial measurement, initial comparison, a proposal plus a revision to the system or model, a new prediction or measurement (depending on which type of revision was made), and finally a new comparison. If the student documented all these steps, they were considered to have completed the modeling process for this activity. Fig. \ref{fig:resactivity2} was organized (left to right) by the number of modeling steps and by the number of proposals plus revisions that were documented, from most to least.

In this activity $53\%$ of students completed the modeling process. On average, students made 2 to 3 proposals to revise the model/system and performed 1 to 2 revisions. 

The initial measurement, proposal, and revision were the most frequently documented components of the modeling process. The initial comparison was the least frequently documented component, as we saw for Activity 1.

The initial prediction of the light intensity was determined prior to class. Despite this, eight of the students did not document it in their notebook. Potentially, they did not recognize the need to document it, since they had already done so in their pre-lab activities. This is further supported by the fact that four of the students who did not document their prediction did document a comparison. These comparisons referenced the pre-lab calculation of their prediction of light intensity and not a prediction made during the lab activity.

The comparison being made was between a quantitative measurement of light intensity (using their photodiode) and the students' prediction. Similar to Activity 1, students were prompted to make proposals and revisions and thus regardless of whether or not they evaluated their comparison to be in good agreement, the students proceeded with revising their model or system. There were two predominant types of comparisons (for both the initial and new comparison) that students performed:
\begin{itemize}
\item{Quantitative: These comparisons were made by either stating a percentage difference, ratio, or order of magnitude difference between the measurement and prediction. Generally, students did not state a threshold for this comparison to distinguish good agreement from poor agreement, and thus it was unclear how these comparisons were used as justification for subsequent revision (\emph{e.g.}, top of Fig. \ref{fig:compact2}).}
\item{Qualitative: These comparisons consisted of a subjective qualitative statement. For example, the student might have stated that the measurement was ``a little less than the prediction.'' It was generally unclear (like the quantitative comparisons) how these comparisons were used as justification for subsequent revisions students made to the model/system.}
\end{itemize}

Examples of these two types of comparisons can be seen in Fig. \ref{fig:compact2}.
\acttwocompex

The proposal and revision steps were the main focus of this activity. Unlike Activity 1, Activity 2 prompted students to make their own choice about how to revise the model or apparatus. Some specific examples of the most common proposed revisions to the system were \emph{block background light}, \emph{remove light bulb diffuser}, and \emph{reposition photodiode and bulb}. Examples of proposed revisions to the model were \emph{account for the actual number of light bulbs present}, \emph{improve accuracy of bulb-to-diode distance}, \emph{adjust assumption of bulb wattage/intensity}, and \emph{adjust assumption of diode sensitivity}. Of all the proposed revisions, $78\%$ were proposed for the model and $22\%$ were proposed for the apparatus. 

Specific examples of the most common revisions students actually performed on the system were \emph{blocking extraneous light bulbs} and \emph{blocking sunlight}. Examples of revisions performed on the model were \emph{updating number of light bulbs}, \emph{updating distance of light bulbs}, and \emph{estimating intensity of an individual light bulb}. Of all the performed revisions, $82\%$ of them were revisions to the model.

An interesting result from Activity 2 was that despite the broad range of proposals made by students for revision, the range of revisions students ultimately performed were limited. Students made a total of 12 different proposals (the most proposed by a single student was seven) and a total of eight different revisions. The two predominant proposals/revisions pertained to the assumption about the number of bulbs generating light and the distance of these bulbs. These two comprised $75\%$ of the revisions performed but only $54\%$ of the proposals. 

These two revisions related to two of the three major initial assumptions students made when making their initial prediction (number and distance of bulbs modeled). The third main assumption pertained to the choice of peak wavelength for the bulbs. Given that students had limited ability to assess the complex spectrum of the fluorescent bulbs, revisions to the model addressing this third assumption would have been the most difficult to incorporate. Thus, despite the students' creativity in proposing revisions, it is likely that the revisions they performed were chosen because they were the easiest and most evident to make.

\subsection{\label{sec:resact3}Activity 3: Voltage-controlled electromagnet}
\actthreecoderes

Activity 3 had the potential for multiple iterations of modeling, with a number of different models to describe the magnetic field. These options consisted of both qualitative and quantitative models. The coding results for Activity 3 can be seen in Fig. \ref{fig:resactivity3}. At minimum, a complete modeling cycle consisted of documentation of an initial prediction, initial measurement, and an initial comparison. The modeling cycle was complete if the students' comparison indicated agreement between the prediction and measurement. Alternatively, if the comparison did not indicate agreement then the student was expected to revise the system. If the student documented all of these steps, they were considered to have completed the modeling process for this activity. Fig. \ref{fig:resactivity3} was organized (left to right) by the number of modeling steps and by the number of proposals plus revisions that were documented.

In this activity, $68\%$ of students completely documented a modeling cycle. Of those, all but two felt their measurement agreed with their prediction and therefore did not proceed with a revision to their model or system. Only these two students documented revisions to their model but neither made subsequent measurements or predictions. Though this activity had a good deal of opportunity for iterative modeling cycles, no student proceeded thusly.

Instead of iterating on a their model, a number of students tested multiple features of their model by taking measurements of different aspects of the B-field (dependence on current, number of turns, distance along z-axis, \emph{etc.}) using different measurement tools (gaussmeter, permanent magnet, compass, \emph{etc.}). Many of these additional measurements resulted in distinct initial comparisons, but did not prime students to perform subsequent modeling cycles, in contrast to the previous two activities. These additional documented measurements and comparisons are denoted in Fig. \ref{fig:resactivity3}: 22 students documented more than one measurement and 12 students documented more than one comparison.

The initial prediction and initial measurement were the most frequently documented components of the modeling process. The least frequently documented component was the revision.

The initial predictions were generated from various models that ranged from quantitative to qualitative. The models students used related to the five suggestions in the lab guide (step D of the lab guide in Sec. \ref{sec:act3}). Examples of the models students tested included the \emph{direction of the B-field on either side of the solenoid}; the \emph{B-field's dependence on current or number of turns in the solenoid}; or the \emph{behavior of the B-field as a function of distance along the axis of the coil}. The most commonly tested model was the axial B-field ($B_z$) as a function of distance from the plane of the solenoid. Students used the equation for $B_z$ along the axis of a current carrying loop and measured the field using the gaussmeter. Other commonly tested models were the orientation of the B-field on either side of the solenoid or the trend of the magnitude of the B-field as a function of distance (in all directions) from the solenoid, both using a compass as their measurement device.

In Activity 3, the specifics of the comparison depended on which model was being tested. Much like the previous activities, there were both qualitative and quantitative types of comparisons. The following were the most common types of comparisons:
\begin{itemize}
\item{Plotting (quantitative): This was the most detailed comparison students made, where they plotted their theoretical model along with their data points. This was only done for students using the axial B-field ($B_z$) of the solenoid as their model and a Gaussmeter as their measurement tool. However, despite the quantitative nature of this comparison, students did not make quantitative evaluations of this comparison.}
\item{Proportionality (quantitative/qualitative): This comparisons consisted of students testing a model describing the proportionality of the B-field as a function of the tested variable. An example of this can be seen in Fig. \ref{fig:compact3} (middle), where the student is testing the model that the B-field is linearly proportional to both the number of turn and the current in the coil.}
\item{Trend (qualitative): This consisted of students comparing the trends of the B-field (increasing or decreasing) to their model as a function of the variables being tested. For example, verifying that the B-field decreased as you moved away from the coil. This type of comparison was the least sophisticated.}
\end{itemize}

Examples of these three types of comparisons can be seen in Fig. \ref{fig:compact3}.
\actthreecompex

Furthermore, of the 36 students who documented their comparisons only seven students indicated poor agreement between their prediction and initial measurement. These seven students are indicated by X's on the corresponding comparisons, in Fig. \ref{fig:resactivity3}. The remaining comparisons were evaluated as demonstrating good or sufficient agreement.

Students made a number of different proposals to revise their model/system (approximately 10 distinct proposals across all students). The most common proposals were to \emph{improve the shape of the solenoid}, by making it flatter or more circular, \emph{stabilize the orientation of the solenoid and measurement device} by using a mount, and \emph{shielding the background magnetic field}. Most were proposals to revise the system. Unlike Activity 2, these proposals were generally not as easy to implement. The students identified many potential reasons for their discrepancy, but generally they did not provide actionable ways of implementing these proposed revisions, given the available equipment. As stated above, seven students documented comparisons that indicated poor agreement, but these students comprised only about half of those who went on to propose revisions.

An interesting result from Activity 3 was that essentially none of the comparisons resulted in students revising their model/system (as seen in the coding results, Fig. \ref{fig:resactivity3}), regardless of whether or not their comparison indicated good agreement. Most students evaluated their comparison to be sufficient enough to stop the modeling process. However, 11 students acknowledged that the model/system could be improved and proposed general ideas for how to do so, but did not proceed with a revision. In some cases, the students' comparison did not aid them in identifying productive ways of revising their model. For example, students might use the equation for $B_z$ along the axis of a current carrying loop for their model and generate a plot at the top of Fig. \ref{fig:compact3}, but then compare their prediction and measurement by making a qualitative statement like ``our data demonstrates the correct trend'' and thereby stopping the modeling process. In other words, students utilized qualitative comparisons for more quantitative models. This allowed them to justify why they stopped the modeling process.

\section{\label{sec:disc}Discussion}

In the discussion, we address three questions: (i) Are students documenting recursive modeling cycles in their notebooks? (ii) Do differing levels of scaffolding in lab activities influence students' documented modeling cycles? If so, how? (iii) With which components of the modeling process do students demonstrate more or less proficiency? Each of the following sections outline our findings for one of these questions.

\subsection{Presence of modeling in notebooks}

In general, students were capable of documenting model-based reasoning in their lab notebooks. The students demonstrated that they were able to follow the activity prompts and use the relevant modeling vocabulary, consistent with how it's used in the lab guide. This allowed us to identify each of the components of the modeling process. Through our coding, we were able to identify that that a majority of students were able to fully document at least one modeling cycle for each of the three activities (74\%, 53\%, and 68\%, for Activities 1--3 respectively), indicating that the goal of the activities was effectively communicated.

On the other hand, the fact that 74\% of students fully documented at least one modeling cycle in Activity 1, but only 37\% completed all four parts of the activity indicates that despite the fact that most students were capable of documenting their process, they were not consistent on the whole.

Activities 2 and 3 consisted of at most two modeling cycles. The bar for what constituted completion of the modeling process for Activity 2 was lower than that for the first activity, and what constituted a complete process was lower still for Activity 3. This fact may be why a greater percentage of students completed the entire modeling process for the latter two activities.

Ultimately, since we were looking at student notebooks, we were capturing only the modeling students actually \emph{took the time to document}. It is likely students were using model-based reasoning in real time to a greater degree than that which they documented. The fact that most students documented at least one modeling cycle for Activity 1, but were not being thorough enough to document the full four parts of the process is indication of this. Requiring students to document their modeling is a higher bar for engagement than simply using real time model-based reasoning. Furthermore, documentation is a skill that students at this stage are still in the process of developing\cite{Stanley2016}, and therefore they may not be as thorough in recording their reasoning. This is one limitation for the approach taken herein. If students do not explicitly record this information, it is unlikely their reasoning could be assessed in a course setting. Therefore additional effort is needed to motivate students to do the additional step of documenting their reasoning if one wants to evaluate model-based reasoning in this manner. An alternative approach is the think-aloud interviews referenced in the introduction. However, these interviews make it difficult to study the modeling of large numbers of students, whereas using students' documentation facilitates a much larger sample size.

\subsection{Effects of scaffolding on modeling}

Scaffolding did appear to have an effect on the students' ability to complete the modeling cycles.

The students demonstrated differing levels of engagement with modeling for differing levels of scaffolding in the activities. For the most scaffolded activity (Activity 1), students documented the process of going through multiple modeling cycles for their two circuits. Generally, all students were able to follow the process of multiple iterations of the modeling cycle. Although only about 37\% completed the full activity (all four parts), it should be noted that what constituted a ``complete'' process for this activity consisted of approximately four times more steps than either of the other two activities. Most of the remaining students missed only one or two steps in the modeling process of Activity 1.

In Activity 2, although they were required to make some revision, students were not prompted to propose and perform specific revisions to the model/system. As a result, students proposed a broad range of differing ideas for how the model/system could be revised in order to improve the agreement between measurement and prediction. However, the spectrum of revisions students ended up performing was much smaller. These were the revisions (mostly made to the model) that were the easiest to implement.

In Activity 3, students were required to test the predictions of their model of choice, but were not explicitly required to perform revisions to the model/system. Instead the choice to revise the model/system hinged on the students' evaluation of their comparisons. As a result, the majority of students made no proposal for revision and only two actually revised their model/system. This was in spite of the fact that a number of students identified a need for revision (and, subsequently, the generation of a new measurement or prediction) in their comparison. It should be noted that this activity was a part of the final lab and thus much of the class was likely concerned with finishing their lab work in the remaining time---indicating agreement in their comparison would allow them to finish the activity more quickly.

Given the lack of scaffolding in the activity prompts, students may have been motivated to adopt a lower bar for what constituted sufficient agreement in their comparison. This would allow them to stop the modeling process sooner, thus saving them spending further time and energy.
 
Ultimately, it is unclear if students' lack of modeling iteration in this case was due to a limited ability to interpret their comparison, limited ability to devise a revision, limited time and resources to devote to continuing the modeling process, or a combination of these. In any case, the lack of scaffolding in the activity corresponded to a decreased engagement with the modeling process in comparison to the previous two activities.
 
\subsection{Proficiency with components of modeling}

Prediction and Measurement: It is clear from all three activities that students could communicate their predictions and measurements. Documentation of students' measurements was nearly universal. For those instances where students were not documenting their predictions, many students still made comparisons and thus we can infer they were aware of their predictions. In the case of the remaining absent predictions, it is unclear whether or not the students were aware of the importance of the prediction for the activity.

Proposal and Revision: When prompted, students demonstrated the ability to propose ideas for revisions to their model/system and carry out a number of them, as demonstrated by their documentation of Activities 1 and 2. When prompted to propose their own revisions students demonstrated a great deal of creativity and novelty in their solutions, but their choice of revision to perform was still constrained by pragmatic factors such as time and equipment availability---the fact that most students chose the same revisions in Activity 2 is an indication of this. However, when students were not explicitly instructed to revise their system they did not take the initiative to do so, as was demonstrated by their performance on Activity 3. It is possible this was due to time/equipment constraints, insufficient motivation for course credit, or lack of recognition of the need to revise their model/system.

Comparison: The aspect of the modeling process with which students had the most difficulty was making and communicating their comparisons. Based on our coding results and examination of the specifics of students' comparisons, their difficulty was threefold: students may not have understood the importance of making a comparison to determine if revision is necessary; their criteria for comparisons were not appropriate or sufficiently sophisticated for the model they were testing; or even if students made a comparison, they might not have recognized the importance of documenting it. It should be noted that students were explicitly instructed to specify the criteria they used for comparison, but generally did not.

For example, Activity 1 was the most straightforward activity and the lab guide instructed students to make a comparison, yet a number of students proceeded with revising their model/system without having documented their comparison. For most of those that did document a comparison, the criteria they used to evaluate the comparison were not well explicated nor sufficiently sophisticated to motivate revision. This was also the case for students' comparisons in Activity 2. 

In Activity 3, most students made comparisons that were related to their predictions and measurements, but these comparisons were not sufficiently sophisticated or refined enough to motivate revisions to their model/system. For example, in the case of those students who measured B-field with the gaussmeter and plotted it against a theoretical curve of Biot-Savart law, none addressed measurement uncertainty or utilized statistical tools to evaluate their comparisons and instead made simple qualitative comparisons that focused on the general trends of the data. Despite the apparent quantitative disagreement between their measurement and model, students did not proceed with any revisions.

It should be noted, however, the course framing and structure did not emphasize students' use of quantitative error analysis, as it was not a learning goal, and thus students were not expected to utilize sophisticated error analysis when making their comparisons. With that said, it was not anticipated that this lack of sophistication in their error analysis would prevent them from continuing the modeling process.

As stated above, students were given agency in choosing what criteria to use to evaluate their comparisons, but based on our findings, it is likely that students' lacked the lab experience or experimental context to effectively exercise this agency. Likely, this resulted in the mismatch seen between the type of model they were evaluating and the comparison they chose, seen in both Activities 2 and 3. A part of developing experimental acumen for any physicist is being able to evaluate when sophisticated criteria for evaluating predictions and data are needed and when it is reasonable to use rough, order-of-magnitude-type criteria. It is common for lab classes to externally impose a criteria to use, instead of providing students the agency of choosing their criteria. In the case of our course, students were provided the agency to chose, but were not experienced enough to make informed decisions about what was appropriate for their model. For many students, this was their first time working in this kind of lab environment, so they had not developed the expertise in how to evaluate their measurements. So in lieu of expertise, the course might have provided students with guidelines for how they should have gone about this step in the modeling process. However, this would deny students the opportunity to learn about and actively engage with differing levels of sophistication for their comparison criteria. This may be a fundamental problem with lab courses that attempt to develop authentic experimental lab skills.

Ultimately, it appears the comparison step in the modeling process was the keystone component for the three activities---if the lab activity did not explicitly instruct students to complete each step in the modeling process then the quality of the students' comparison would dictate whether or not they continued iterating on the activity. Therefore, understanding the nature of students' comparisons could help to make sense of their approach to the modeling process as a whole.

\section{\label{sec:conc}Conclusions}

This work has provided a foundation for understanding how to evaluate model-based reasoning through documentation. We have demonstrated how the Modeling Framework for Experimental Physics can be used to examine and assess modeling in students' lab notebook entries recorded during lab activities. Using the EMF in this manner is appropriate for physics lab courses that emphasize engaging students in modeling. We were able to track students reasoning through multiple iterations of modeling in several lab activities with differing levels of scaffolding. We determined the variation in scaffolding across the activities had an impact on how thoroughly students engaged in documenting full modeling cycles. Furthermore, students demonstrated varying degrees of proficiency with the different parts of the modeling process, with the most difficult being the evaluation and communication of their comparisons between the model prediction and the interpreted data from their measurement.

Our findings have implications for how to improve future iterations of the transformed course on which we have focused. Specifically, our results indicate a need to provide more scaffolding for how students should perform the individual components of the modeling process. The component of the modeling process that was the most difficult for students was the comparisons phase. In many lab classes, students are provided with criteria for how to evaluate comparisons they make in their lab activities. These can include rough criteria such as a trend or order of magnitude comparison, or can include more mathematically sophisticated criteria such as a t-test or evaluation of chi-square. These externally imposed criteria provide students with easy to follow rules, which can lighten the cognitive load required for their lab activities. However, the downside to this approach is that students are denied the agency to develop the expertise in how to evaluate comparisons. In the course we have studied, students were not provided such criteria and thus were required to chose their own. However, the students lack of experience with choosing criteria prevented students from continuing with the modeling process.

In the future, we would like to focus on ways of still providing students the freedom to chose the criteria they use to evaluate their comparisons, but also provide support for understanding the range of sophistication for these criteria and how to chose what is appropriate for their specific comparison. Providing students with this kind of scaffolding can help to facilitate their practice with the modeling process more broadly.

\begin{acknowledgments}
We would like to acknowledge and thank Robert Hobbes and Dimitri Dounas-Frazer for their helpful feedback and initial help with coding. This work was supported by NSF grant nos. DUE-1323101 and PHY-1125844.
\end{acknowledgments}
%
%

\end{document}